\documentclass[aip,apl,reprint,amsmath,floatfix,noeprint]{revtex4-2}

\usepackage{graphics}
\usepackage{graphicx}
\usepackage{epsfig}
\usepackage{amsmath}
\usepackage{bm}
\usepackage{lineno}
\usepackage{color}
\usepackage{dcolumn}
\usepackage[utf8]{inputenc}
\usepackage[T1]{fontenc}
\usepackage{mathptmx}
\begin{document}
\preprint{AIP/123-QED}
\title{Two-magnon frequency-pulling effect in ferromagnetic resonance}
%
\author{W.~K.~Peria}
\affiliation{School of Physics and Astronomy, University of Minnesota, Minneapolis, Minnesota 55455, USA}
\author{H.~Yu}
\affiliation{Department of Materials Science and Engineering, University of Maryland, College Park, Maryland 20742, USA}
\author{S.~Lee}
\affiliation{Department of Materials Science and Engineering, University of Maryland, College Park, Maryland 20742, USA}
\affiliation{Department of Physics, Pukyong National University, Busan 48513, South Korea}
\author{I.~Takeuchi}
\affiliation{Department of Materials Science and Engineering, University of Maryland, College Park, Maryland 20742, USA}
\author{P.~A.~Crowell}
\email[Author to whom correspondence should be addressed: ]{crowell@umn.edu}
\affiliation{School of Physics and Astronomy, University of Minnesota, Minneapolis, Minnesota 55455, USA}
\date{\today}
\begin{abstract}
We report the experimental observation in thin films of the hybridization of the uniform ferromagnetic resonance mode with nonuniform magnons as a result of the two-magnon scattering mechanism, leading to a frequency-pulling effect on the ferromagnetic resonance. This effect, when not properly accounted for, leads to a discrepancy in the dependence of the ferromagnetic resonance field on frequency for different field orientations. The frequency-pulling effect is the complement of the broadening of the ferromagnetic resonance lineshape by two-magnon scattering and can be calculated using the same parameters. By accounting for the two-magnon frequency shifts through these means, consistency is achieved in fitting data from in-plane and perpendicular-to-plane resonance conditions.
\end{abstract}
\maketitle
Magnetization dynamics of ferromagnets are influenced by many factors, such as magnetocrystalline anisotropy, interfacial anisotropy, and Gilbert damping. It is of technological interest to study these phenomena in order to understand, for example, the physics governing magnetization switching by spin torque.\cite{Ralph2008,Sinova2015,Manchon2019} Interfacial anisotropies are particularly relevant for these applications, where materials with perpendicular anisotropy are sought due to the lower energy cost associated with switching of the magnetization.\cite{Hellman2017} Anisotropy arising from broken symmetry at the interfaces in ultrathin magnetic films is commonly probed using ferromagnetic resonance (FMR) techniques,\cite{Beaujour2009,Chen2018} usually by measuring the FMR fields as a function of driving frequency or applied field orientation.

Two-magnon scattering (TMS), an extrinsic relaxation process of uniform magnetization precession in ferromagnets, is an important phenomenon that influences magnetization dynamics. TMS is commonly observed as a broadening of FMR lineshapes, but it may also lead to a shift in the FMR frequency.\cite{Arias1999,McMichael2004,Krivosik2007} Although the broadening of the FMR lineshape caused by TMS is often impossible to ignore, the latter effect is more subtle and almost universally neglected when FMR data are used to extract static magnetic properties of materials. Failure to properly account for this effect, however, may lead to inaccurate estimates of magnetic anisotropy energies, potentially leading to the inference of a substantial interface anisotropy where there is none.

In this letter, we demonstrate the existence of a frequency-pulling effect in ferromagnetic resonance induced by two-magnon scattering in polycrystalline Fe$_{0.7}$Ga$_{0.3}$ thin films for in-plane applied magnetic fields. We arrive at this result by calculating the frequency shift based on our analysis of the linewidth broadening caused by two-magnon scattering (which is a large effect and is easier to observe). This is possible due to the complementary nature of the two phenomena. We show that the observed resonance frequencies are only consistent with the data taken for perpendicular-to-plane fields (i.e., can be fit using the same perpendicular anisotropy field) when they are adjusted to account for the two-magnon frequency shifts. We conclude by demonstrating this effect in additional samples of different thicknesses, simultaneously showing that the magnitude of the frequency shifts scales with the magnitude of the two-magnon linewidths as the theory predicts.

\begin{figure}
  \centering
  \includegraphics{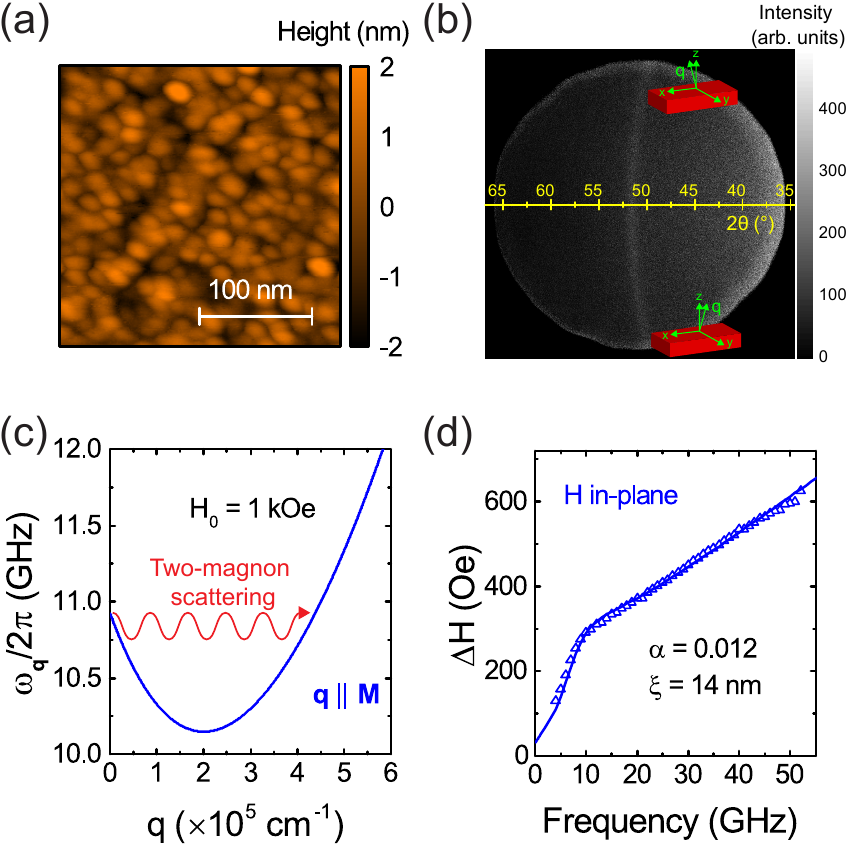}
  \caption{(a)~Atomic force microscopy image of the 33~nm Fe$_{0.7}$Ga$_{0.3}$ film. Root-mean square roughness is 0.7~nm. (b)~Two-dimensional XRD detector image of the 33~nm Fe$_{0.7}$Ga$_{0.3}$ film showing the (110) Bragg peak (given by the ``ring'' at $2\theta \simeq 52^\circ$). The center of the detector corresponds to the symmetric configuration, in which the scattering vector $\mathbf{q}$ is normal to the plane of the film ($q_x = q_y = 0$). The scattering vector is canted into the film plane as one moves vertically from the center of the detector, as indicated by the coordinate axes. The structural coherence length determined from the full-width-at-half-maximum of the Bragg peak is 13~nm. (c)~Thin-film magnon dispersion for in-plane magnetization and wavevectors $\mathbf{q} \parallel \mathbf{M}$, with an arrow indicating the two-magnon scattering process. (d)~Field-swept FMR linewidths of the 33~nm Fe$_{0.7}$Ga$_{0.3}$ film with in-plane applied magnetic field overlaid with a fit to a combined two-magnon scattering and Gilbert damping model. The Gilbert damping $\alpha$ (a fit parameter) and defect correlation length $\xi$ (fixed) are shown on the figure.}\label{fig:fig1}
\end{figure}

The samples used in this report are 17~nm, 26~nm, and 33~nm Fe$_{0.7}$Ga$_{0.3}$ films (thicknesses determined by x-ray reflectivity). The 33~nm films were deposited on Si/SiO$_2$ substrates at room temperature by dc magnetron sputtering of an Fe$_{0.7}$Ga$_{0.3}$ target. The base pressure of the deposition chamber was $5 \times 10^{-8}$~torr and the working pressure was maintained at $5 \times 10^{-3}$~torr by Ar gas (99.999\%). The composition of the Fe$_{0.7}$Ga$_{0.3}$ films was quantitatively analyzed by energy dispersive spectroscopy (EDS). The 17~nm and 26~nm films were obtained by etching the 33~nm films with an ion mill. The lack of magnetic anisotropy in the plane of the film was verified with vibrating sample magnetometry (VSM) and FMR for the 33~nm film. Grain boundaries were observed with atomic force microscopy (AFM), yielding an average grain diameter of $\sim$15~nm [see Fig.\ \ref{fig:fig1}(a)]. This is in good agreement with the structural coherence length, which was estimated to be 13~nm with XRD. Figure \ref{fig:fig1}(b) shows a two-dimensional XRD detector image of the Fe$_{0.7}$Ga$_{0.3}$ (110) peak for the 33~nm film, where the center of the detector coincides with a scattering vector normal to the film plane. The intensity of this Bragg peak is approximately constant for fixed values of the scattering angle $2\theta$ as the scattering vector is canted into the plane [as evidenced by the ``ring'' in Fig.\ \ref{fig:fig1}(b)], indicating the absence of texture.

Ferromagnetic resonance lineshapes were measured at room temperature using a coplanar waveguide with modulation of the applied magnetic field for lock-in detection as described in Ref.\ \onlinecite{Peria2020}. The coplanar waveguide was placed in series with a rectifying diode that measured the transmitted microwave power. The applied dc magnetic field was modulated with a 220~Hz ac magnetic field having an amplitude of a few Oe for lock-in detection of the differential absorption. The microwave frequency was varied up to 52~GHz with power from 5 to 10~dBm. The lineshapes were measured for both in-plane (IP) and perpendicular-to-plane (PP) applied fields. When the magnetization is IP, there exist magnons degenerate with the $\mathbf{q} = 0$ FMR magnon [see Fig.\ \ref{fig:fig1}(c)]. This leads to a possible scattering mechanism of the FMR mode, observable through its nonlinear effect on the field-swept linewidth as a function of frequency,\cite{Arias1999,McMichael2004,Krivosik2007} shown in Fig.\ \ref{fig:fig1}(d) for the 33~nm Fe$_{0.7}$Ga$_{0.3}$ film.  This is the TMS process, and it is allowed as long as some assisting process enables conservation of momentum.

The resonance frequency was fit as a function of the applied magnetic field $H_0$ to the Kittel equation for a thin film with no in-plane magnetic anisotropy, which reads as
\begin{equation}\label{eq:KittelIP}
  \omega_{FMR} = \gamma \sqrt{H_0 (H_0 + 4 \pi M_{eff})}
\end{equation}
for $H_0$ in the plane and
\begin{equation}\label{eq:KittelPP}
  \omega_{FMR} = \gamma (H_0 - 4 \pi M_{eff})
\end{equation}
for $H_0$ perpendicular to the plane,\cite{Farle1998} with $\gamma$ the gyromagnetic ratio and $4 \pi M_{eff}$ the effective demagnetizing field. Henceforward we will express the gyromagnetic ratio in terms of the Land\'e $g$-factor, \textit{via} $\gamma = g \mu_B / \hbar$.
\begin{figure}
  \centering
  \includegraphics{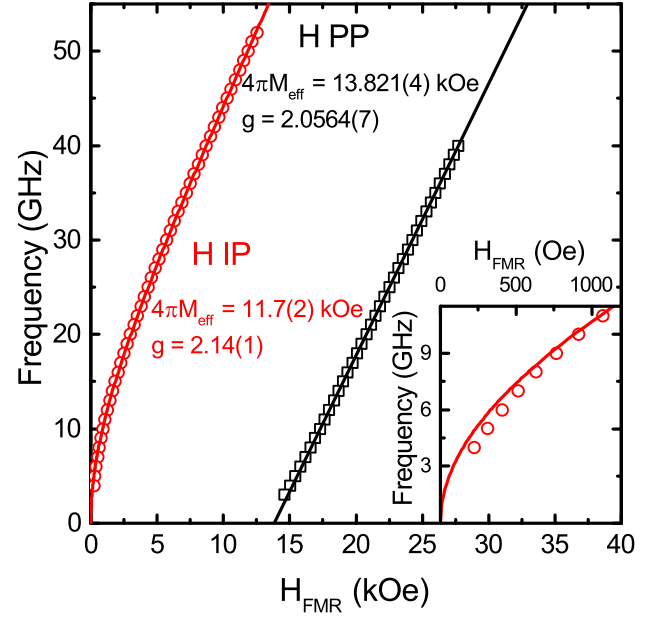}
  \caption{Frequency as a function of resonance field for the 33~nm film obtained with IP (red circles) and PP (black squares) orientations of the magnetic field, overlaid with fits to Eqs.\ (\ref{eq:KittelIP}) and (\ref{eq:KittelPP}), respectively. Fit parameters $4 \pi M_{eff}$ and $g$-factor are indicated on the figure for both cases. Inset shows a close-up of the IP orientation at low frequencies.}\label{fig:bothdispersions}
\end{figure}
Figure \ref{fig:bothdispersions} shows field-dependent resonance dispersions of the 33~nm film for IP and PP applied fields, along with fits to Eqs.\ (\ref{eq:KittelIP}) and (\ref{eq:KittelPP}). For the PP case we obtain $g = 2.0564 \pm 0.0007$ and $4 \pi M_{eff} = 13.821 \pm 0.004$~kOe, and for the IP case $g = 2.14 \pm 0.01$ and $4 \pi M_{eff} = 11.7 \pm 0.2$~kOe. The PP value of $4 \pi M_{eff} = 13.821$~kOe is higher than previous bulk measurements \cite{Clark2005} but is similar to values obtained in thin films. \cite{Gopman2017} The inset shows a close-up of the IP data at low frequencies, whence it is clear that a discrepancy exists between the IP data and the fit to Eq.\ (\ref{eq:KittelIP}). For the 26~nm and 17~nm thicknesses we observe the same qualitative behavior. In the case of the 26~nm film we measure $4\pi M_{eff} = 14.0736 \pm 0.0006$~kOe and $g = 2.060 \pm 0.001$ for PP fields, compared to $4 \pi M_{eff} = 12.17 \pm 0.01$~kOe and $g = 2.133 \pm 0.007$ for IP fields. For the 17~nm film we measure $4\pi M_{eff} = 13.0049 \pm 0.0005$~kOe and $g = 2.054 \pm 0.001$ for PP fields, compared to $4 \pi M_{eff} = 11.623 \pm 0.006$~kOe and $g = 2.120 \pm 0.004$ for IP fields. In addition, the IP data for both the 17~nm and 26~nm films cannot be fit well to Eq.\ (\ref{eq:KittelIP}) at low fields (similar to what is seen for the 33~nm film, shown in the inset of Fig.\ \ref{fig:bothdispersions}). There is no in-plane magnetic anisotropy, so the discrepancy cannot be attributed to an in-plane anisotropy field. Furthermore, the parameters yielded by the fits in either case are drastically different. In light of these inconsistencies, we proceed to investigate the effect of TMS on the IP field-dependent resonance dispersion.

One of the primary characteristics of TMS is that it can be suppressed by orienting the magnetization perpendicular-to-plane, a result of the disappearance of degeneracies in the spin wave dispersion as the magnetization is rotated perpendicular to the plane.\cite{Hurben1998} Later this fact will be used to control for TMS, allowing the observation of the noninteracting or ``bare'' properties when the film is perpendicularly magnetized.

The breaking of momentum conservation in TMS necessitates the presence of defects in order to drive the process. There are numerous categories of defects which may cause TMS. Among the prominent ones are surface roughness,\cite{Arias1999,Dobin2004} dislocation networks,\cite{Woltersdorf2004} and grain boundaries.\cite{McMichael2004,Krivosik2007} We will focus here on TMS induced by grain boundaries, having confirmed the structural isotropy of the films with XRD as well as observing grains directly with AFM. These characterization data also allow us to constrain the defect lengthscale, which partially determines the strength of coupling between $\mathbf{q} = 0$ and $\mathbf{q} \neq 0$ magnons. In the context of TMS, the grains lead to a spatially inhomogeneous and random anisotropy field. The inhomogeneous field allows for interaction between the FMR mode and modes at nonzero wavevector, providing both an additional relaxation channel and a change in the effective stiffness of the FMR mode.  These can be described as imaginary and real shifts in the FMR frequency, respectively. A perturbative model of TMS for this system gives the following result for the complex frequency shift of the FMR mode due to interactions with modes at nonzero wavevector:\cite{McMichael2004,Krivosik2007,Kalarickal2008}
\begin{widetext}
\begin{equation}\label{eq:scattrate}
  \Delta \omega_{TMS} = \gamma^2 \xi^2 H'^2 \int d^2 \mathbf{q} ~ \Lambda_{0 \mathbf{q}} \frac{1}{(1 + (q \xi)^2)^{3/2}} \frac{1}{\pi} \frac{1}{(\omega - \omega_{FMR}) - i \delta \omega}
\end{equation}
\end{widetext}
where $H'$ is the root-mean square inhomogeneity field, $\xi$ is the defect correlation length, $\Lambda_{0 \mathbf{q}}$ is the magnon-magnon coupling strength (see Ref.\ \onlinecite{Kalarickal2008}), and $\delta \omega = (d\omega/dH|_{H_{FMR}}) (\alpha \omega / \gamma)$ is the Gilbert frequency half-width-at-half-maximum linewidth. The imaginary part of Eq.\ (\ref{eq:scattrate}) corresponds to the well-known TMS contribution to the FMR scattering rate, i.\,e.~linewidth. Lesser known, however, is the real part, which describes a shift of the FMR frequency due to TMS. This effect has been previously reported in ultrathin films of Ni$_{0.5}$Fe$_{0.5}$,\cite{Azevedo2000} but a lack of broadband measurements leaves the results open to interpretation (such as the possibility of it having arisen from interface anisotropy). In addition, the strength of two-magnon scattering in our system is much greater.

We begin our analysis by fitting the field-swept linewidths to the imaginary part of Eq.\ (\ref{eq:scattrate}), including contributions from Gilbert damping (linear with frequency) and inhomogeneous broadening (constant with frequency, determined from the PP measurement). We hold the defect correlation length $\xi$ fixed to 14~nm based on the structural characterization described earlier. The fit parameters are $\alpha$ and $H'$, which we use to calculate the FMR frequency shifts from the real part of Eq.\ (\ref{eq:scattrate}). Notably, the Gilbert damping $\alpha$ determines both the Gilbert linewidth \textit{and} the two-magnon linewidth---the latter is clear upon inspection of Eq.\ (\ref{eq:scattrate}) and is discussed at length in Ref.\ \onlinecite{Peria2020}.

\begin{figure}
  \centering
  \includegraphics{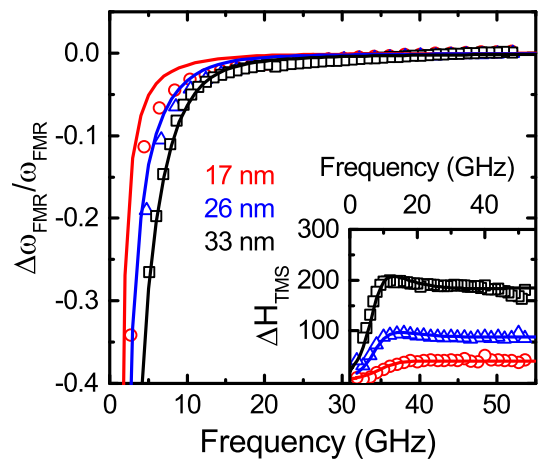}
  \caption{Frequency shifts induced by two-magnon scattering for the 17~nm (red points), 26~nm (blue points), and 33~nm (black points) Fe$_{0.7}$Ga$_{0.3}$ films. The predicted frequency shifts given by the solid curves are calculated from the real part of Eq.\ (\ref{eq:scattrate}) using the fit parameters from the fits of the linewidths. Inset shows the two-magnon linewidths for the three films along with fits to the imaginary part of Eq.\ (\ref{eq:scattrate}). The two-magnon linewidths are determined by subtracting the Gilbert damping and inhomogeneous linewidths (the inhomogeneous linewidths are determined from the PP measurement).}\label{fig:doubleY}
\end{figure}

The fractional FMR frequency shifts for the 17~nm, 26~nm, and 33~nm films are shown in Fig.\ \ref{fig:doubleY}. The solid curves give the predicted fractional frequency shifts based on the fits of the linewidths. The points in Fig.\ \ref{fig:doubleY} represent the observed fractional frequency shifts, determined by taking the difference between the observed FMR frequencies and the FMR frequencies predicted by Eq.\ (\ref{eq:KittelIP}) (taking $4\pi M_{eff}$ from the PP measurement). The strong frequency-pulling effect of TMS is evident from the main panel of Fig.\ \ref{fig:doubleY}, with red shifts approaching 1~GHz at low frequencies.

The two-magnon linewidths for the three films are shown in the inset of Fig.\ \ref{fig:doubleY}. The solid curves give the fits to the imaginary part of Eq.\ (\ref{eq:scattrate}), while the points are obtained by taking the observed linewidths and subtracting both the Gilbert and inhomogeneous broadening (taken from the PP measurement) contributions.
A notable observation from Fig.\ \ref{fig:doubleY} is the correlation between the magnitudes of the two-magnon linewidths and frequency shifts for the three films, which is a prediction of Eq.\ \ref{eq:scattrate}.

\begin{figure}
  \centering
  \includegraphics{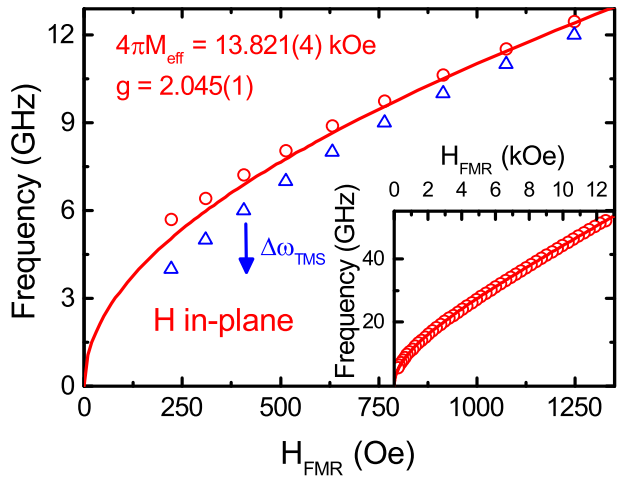}
  \caption{Ferromagnetic resonance frequencies of the 33~nm film at low fields for the IP configuration. The red data points are blue-shifted by amounts given by the red curve in Fig.\ \ref{fig:doubleY}, overlaid with a fit to Eq.\ (\ref{eq:KittelIP}). The effective demagnetizing field $4 \pi M_{eff}$ is fixed to 13.821~kOe based on the fit to the PP data (black squares of Fig.\ \ref{fig:bothdispersions}); the only fit parameter is the Land\'e $g$-factor. The blue data points are the observed resonance frequencies, before two-magnon interactions are taken into account. The inset shows all of the adjusted resonance frequencies up to high fields.}\label{fig:shifted}
\end{figure}

We now discuss how the aforementioned inconsistency between IP and PP field-dependent resonance dispersions, and the inability to obtain a good fit of the IP data to Eq.\ (\ref{eq:KittelIP}), can be explained by the frequency-pulling effect of TMS. The absence of TMS for PP magnetization is of particular convenience in our case because it allows determination of the effective demagnetizing field $4 \pi M_{eff}$ and the Land\'e $g$-factor, which together can in principle be used to predict the FMR field-dependent dispersion for IP magnetization. A direct measurement of the dispersion for IP magnetization is not possible due to the frequency shifts caused by TMS. To address this problem, the IP FMR frequencies are blue-shifted using the red curve shown in Fig.\ \ref{fig:doubleY}, representing the FMR frequencies in the absence of TMS. We then fit the corrected IP FMR frequencies to Eq.\ (\ref{eq:KittelIP}), fixing $4 \pi M_{eff}$ to the PP value of 13.821~kOe (for the 33~nm film) and leaving the $g$-factor as a free parameter. (A small amount of surface anisotropy may lead to an anisotropy of the orbital moment of the film, leading to an anisotropic $g$-factor.\cite{Farle1998}) The shifted IP FMR frequencies at low fields, along with a fit to Eq.\ (\ref{eq:KittelIP}) for fixed $4 \pi M_{eff}$, are shown in Fig.\ \ref{fig:shifted} for the 33~nm film---these values represent the bare FMR frequencies in the absence of two-magnon scattering. The fit yields $g = 2.045 \pm 0.001$, which is less than 1\% lower than the PP value of $g = 2.0564 \pm 0.0007$. Also shown are the FMR frequencies before being adjusted for two-magnon interactions (blue data points). The inset of Fig.\ \ref{fig:shifted} shows the bare FMR frequencies and fit up to high fields. This process was also carried out for the 26~nm and 17~nm films, whereby the agreement with Eq.\ (\ref{eq:KittelIP}) was significantly improved. The fits of the IP data to Eq.\ (\ref{eq:KittelIP}) yielded $g = 2.039 \pm 0.002$ for the 26~nm film and $g = 2.045 \pm 0.002$ for the 17~nm film (both $\lesssim 1 \%$ lower than the PP values).
It is clear from Fig.\ \ref{fig:shifted} that the frequency-pulling effect of TMS is successful at explaining the inconsistencies we have encountered.

In conclusion, we observe a frequency-pulling effect of the ferromagnetic resonance in thin films of Fe$_{0.7}$Ga$_{0.3}$ for magnetization in the plane of the film. It is shown that this effect can be explained by the hybridization of the ferromagnetic resonance with nonuniform magnons as a result of the two-magnon scattering interaction. The frequency shifts can be predicted from the two-magnon induced broadening of the lineshapes, whereby a consistency is obtained with the lineshapes measured when the magnetization is perpendicular to the plane of the film. Our results highlight the importance of accounting for two-magnon scattering when using ferromagnetic resonance as a characterization technique, a fact which is usually ignored in the determination of static properties.

This work was supported by SMART, a center funded by nCORE, a Semiconductor Research Corporation program sponsored by NIST. Parts of this work were carried out in the Characterization Facility, University of Minnesota, which receives partial support from NSF through the MRSEC program, and the Minnesota Nano Center, which is supported by NSF through the National Nano Coordinated Network, Award Number NNCI - 1542202.

The data that support the findings of this study are available from the corresponding author upon reasonable request.
%

%
\end{document}